\documentclass[conference]{IEEEtran}
\usepackage{amsmath,amssymb,amsthm}
\usepackage{dblfloatfix}
\usepackage[english]{babel}
\usepackage{graphicx}
\usepackage{cite}

\newtheorem{theorem}{Theorem}

\newtheorem{lemma}{Lemma}

\newtheorem{definition}{Definition}
\newtheorem{remark}{Remark}

\begin{document}
\title{Resilient Feedback Controller Design For Linear Model of Power Grids}
\author{\IEEEauthorblockN{Amirsina Torfi}
\IEEEauthorblockA{Virginia Polytechnic Institute and State University \\
Department of Computer Science\\
Email: amirsina.torfi@gmail.com
}}

\maketitle
\begin{abstract}

In this paper, a resilient controller is designed for the linear time invariant (LTI) systems subject to attacks on the sensors and the actuators. A novel probabilistic attack model is proposed to capture vulnerabilities of the communication links from sensors to the controller and from the controller to actuators. The observer and the controller formulation under the attack is derived. Thereafter, By leveraging Lyapunov functional methods, it is shown that exponential mean square stability of the system under the output feedback controller is guaranteed if a certain LMI is feasible. The simulation results show the effectiveness and applicability of the proposed controller design approach.  

\end{abstract}

\section{Introduction}

With the advent of digital network technology, the networked control systems (NCSs) have found successful application in industrial processes \cite{keliris2016}, power grids \cite{khodaparastan20121} and so forth because of their important advantages such as low cost, improved utilizing of resources, and simplicity of maintenance. But the vulnerability of communication links to the cyber attacks in NCS has increased. Some factors such as the adoption of open communication standards and protocols, novel energy storage technologies \cite{super,khodaparastan2017study} and connection to the internet help the attackers to launch a successful attack on different NCSs. To deal with security concerns in the modern control systems, it is essential to make the control systems resilient to the malicious attackers. 

The topic of the resiliency of control systems to different types of cyber attacks is considered in several research on designing cyber-resilient systems \cite{ji2011resilient,rieger2012agent, giorgi2012adaptive}. In summary, a resilient control system is determined by its level of resiliency of the system under attacks condition \cite{rieger2010notional}. For instance, the false data injection attacks and the data alteration attacks are examples of the unexpected threats which are used to deteriorate the normal operation of control systems \cite{rieger2009resilient,wei2010resilient,salehghaffari2018}. Also, different approaches such as hybrid analysis \cite{khodaparastan2017novel} and machine learning methods \cite{khodaparastan2012} are utilized for the resilient detection and control in various control systems.  

In recent years, many studies have discussed the definition, mathematical frameworks, and the application of resilient cyber systems in the modern NCS.\cite{zhu2012dynamic} has proposed a game theoretic approach to solve the cascade failure problem in industrial control systems. Thereafter, the interdependency between cyber security and robust control design has been investigated, and some optimal criteria for the linear quadratic case have been obtained. In \cite{hossein2017}, A cyber-resilient controller is proposed for a class of nonlinear discrete systems under actuator attacks where the interaction of the attacker and the IDS is captured with a game in the detection layer. In \cite{elbsat2013robust}, the condition for the existence of a robust resilient state feedback controller for a certain class of nonlinear discrete systems with norm bounded nonlinearity, is obtained. \cite{li2015h} proposes a robust estimator for a class of nonlinear time varying systems with randomly occurring uncertainties, and the specific $H_\infty$ performance is obtained for the estimator by leveraging stochastic stability analysis. In \cite{melin2013mathematical}, a mathematical formulation for the examination of the security problems in the control system has been introduced, and the definition of the several key words in resilient control system design such as state awareness and operational normalcy in a mathematical framework is developed. \cite{pasqualetti2007distributed} considers the problem of malicious behavior of the attackers in the control systems as an unknown input observer design problem. \cite{mo2009secure} has focused on effects of the specific type of attacks in industrial control systems. In \cite{mo2009secure}, the impacts of the reply attacks on the performance of the control systems in the steady state conditions are analyzed. \cite{mo2009secure} shows that the attack is undetectable due to recording the sensors' data and propagating the sensors' data during false data injection attacks on the actuators of the system.\cite{bezzo2014attack} and \cite{fawzi2014secure} analyze the problem of the secure estimation and the secure control of a system when the sensors and the actuators are under attacks.  \cite{fawzi2014secure} shows that it is impossible to recover the states of a system if more than half of the sensors are under the attacks. 

In this paper, a probabilistic attack model is proposed to examine attacks on the sensors and the actuators of the LTI systems. The unreliability of the communication links between sensors to controller and controller to actuators have been modeled by Bernoulli random variables.The fraction of the time that each sensor/actuator is under the attack is modeled by a Bernouli variable. Then, the observer and the controller equations are derived based on expected value of the random variable. Thereafter, the controller is designed based on feasibility of a certain LMI condition obtained by leveraging the stochastic Lyapunov function approach.

This paper is organized as follows: In section~\ref{s2}, the mathematical model of attacks and the closed-loop dynamics of the system under attacks are introduced. The controller design procedure and the stochastic stability analysis are examined in section ~\ref{s3}. The efficiency of the proposed resilient controller is shown by an example in section ~\ref{s4} and conclusions are made at the end.

---------------------------------------------------------------------
\section{Problem Definition} \label{s2}

\begin{figure}[!t]
\centering
\includegraphics[width=0.40\textwidth]{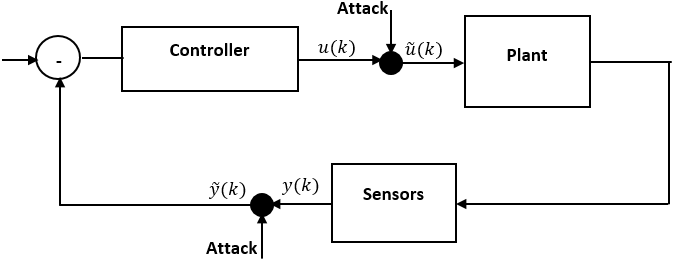}
\caption{The schematic of the control system with unreliable communication links.}
\label{fig1}
\end{figure}

Consider the following discrete time-invariant system:
\begin{eqnarray}
x(k+1)&=&Ax(k)+Bu(k) \label{eq1} \\
y(k)&=&Cx(k)
\end{eqnarray}
where $x(k) \in R^n$ is the system state, $u(k) \in R^m$ is the input control signal, $y(k) \in R^p$ is the measured output, $A$, $B$ and $C$ are the constant matrices with appropriate dimensions. Shown in figure \ref{fig1}, it is assumed the communication links between sensors to controller and controller to the actuators are not secure and the attackers is able to manipulate the sensor and actuator values. 

The proposed attack model for the sensor measurements based on stochastic control theory is given as:
\begin{equation}
\tilde{y}(k)=\Pi_1 Cx(k)+(I-\Pi_1)\Pi_2 C x(k)
\label{eq2}
\end{equation} 
Where $\tilde{y}(k) \in R^p$ is the measured output under attack, $\Pi_1=diag\{\alpha_1, \alpha_2, \ldots, \alpha_p \}$ with $\alpha_i (i=1 , 2, \ldots, p)$ are $p$ uncorrolated stochastic variables coming from a Bernoulli distribution, and $\Pi_2=diag\{\beta_1, \beta_2, \ldots, \beta_p \}$ with $\beta_i (i=1 , 2, \ldots, p)$ are $p$ uncorrelated stochastic variables coming from an unknown probability distribution. The mathematical expectation and variance of the random variables $\alpha_i$ and $\beta_i$ are defined as following:
\begin{eqnarray}
E\{\alpha_i\}&=&Prob(\alpha_i=1)=\bar{\alpha}_i \\
Var\{\alpha_i\}&=&\alpha_{1i}^2 \\
E\{\beta_i\}&=&\bar{\beta_i} \\
Var\{\beta_i\}&=&\beta_{1i}^2 
\end{eqnarray}
where $E(x)$ and $Var(x)$ denote the mathematical expectation and variance of random variable $x$ respectively. $\bar{\beta_i}$ is the expected value of attack injected on the $i^{th}$ sensor, and $\beta_{1i}$ models the deviation from the optimal strategy.
\begin{remark}
The random variable $\alpha_i$ taking value on $\{0,1\}$ has been introduced to represent the attack on the $i^{th}$ sensor. When $\alpha_i$ is equal zero, the correct value of the $i^{th}$ sensor is substituted by the injected attack value. The mathematical expectation of Bernoulli random variable $\alpha_i$ determines the fraction of time that the $i^{th}$ sensor is under attack.
\end{remark}

Given (\ref{eq1}) and (\ref{eq2}), the observer dynamics and the controller attack model are:
\begin{eqnarray}
\hat{x}(k+1)&=&A\hat{x}(k+1)+Bu(k)-L[\tilde{y}-\bar{\Pi}_1C\hat{x}(k) \nonumber \\
&-&(I-\bar{\Pi}_1)\bar{\Pi}_2 C_1\hat{x}(k)] \\
u(k)&=&\Pi_3 K\hat{x}(k)+(I-\Pi_3)\Pi_4 K \hat{x}(k) \label{eq4}
\end{eqnarray}

where $\hat{x}(k) \in R^n$ denotes the system state estimation vector, $L \in R^{n\times p}$ and $K \in R^{m\times n}$ are observer gain and controller gain, $\Pi_3=diag\{\gamma_1, \gamma_2, \ldots, \gamma_m \}$ with $\gamma_i (i=1 , 2, \ldots, m)$ are $m$ uncorrolated stochastic variables coming from a Bernoulli distribution, and $\Pi_4=diag\{\delta_1, \delta_2, \ldots, \delta_m \}$ with $\delta_i (i=1 , 2, \ldots, m)$ are $m$ uncorrelated stochastic variables coming from an unknown probability distribution. The mathematical expectation and variance of the random variables $\gamma_i$ and $\delta_i$ are defined as follows:
\begin{eqnarray}
E\{\gamma_i\}&=&Prob(\gamma_i=1)=\bar{\gamma}_i \\
Var\{\gamma_i\}&=&\gamma_{1i}^2 \\
E\{\delta_i\}&=&\bar{\delta_i} \\
Var\{\delta_i\}&=&\delta_{1i}^2. 
\end{eqnarray}
$\bar{\gamma_i}$ is the expected value of attack injected on the $i^{th}$ actuator, and $\gamma_{1i}$ models the deviation from the optimal strategy.
Since $\Pi_1$, $\Pi_2$, $\Pi_3$, and $\Pi_4$ are diagonal matrices with random independent elements, the following equations can be used to obtain expected value and variance of $\Pi_i\Pi_j (i,j=1,2,3,4)$:
\begin{eqnarray}
E(XY)&=&E(X)E(Y) \\
Var(XY)&=&[E(X)]^2 Var(Y)+[E(Y)]^2 Var(X) \nonumber \\
&+&Var(X)Var(Y)
\end{eqnarray}
where $X$ and $Y$ are two independent random variables. 

The estimation error is defined as:
\begin{equation}
e_k=x_k-\hat{x}_k.
\label{eq5}
\end{equation}

The closed-loop dynamics of the system is obtained By substituting (\ref{eq2}) and (\ref{eq4}) into (\ref{eq1}) and (\ref{eq5}):
\begin{equation} \label{eq51}
\begin{aligned} 
{}&x(k+1)=[A+ B(\bar{\Pi}_3+(I-\bar{\Pi}_3)\bar{\Pi}_4 )K]x(k)+                                                                         \\
&B[(\Pi_3-\bar{\Pi}_3)+(\Pi_4-\bar{\Pi}_4)-(\Pi_3\Pi_4-\bar{\Pi}_3\bar{\Pi}_4)]Kx(k)                                            \\
&-B[\bar{\Pi}_3+(I-\bar{\Pi}_3)\bar{\Pi}_4 ]Ke(k)-B[(\Pi_3-\bar{\Pi}_3)+(\Pi_4-\bar{\Pi}_4)                                  \\
&-(\Pi_3\Pi_4-\bar{\Pi}_3\bar{\Pi}_4)]Ke(k) 
\end{aligned}
\end{equation}
\begin{equation}
\begin{aligned} 
{}&e(k+1)= (A-L(\bar{\Pi}_1+\bar{\Pi}_2-\bar{\Pi}_1\bar{\Pi}_2 )Ce(k)    \\
&-L[(\Pi_1-\bar{\Pi}_1)+(\Pi_2-\bar{\Pi}_2)-(\Pi_1\Pi_2-\bar{\Pi}_1\bar{\Pi}_2)]Cx(k)
\label{eq6}
\end{aligned}
\end{equation}
where $\bar{\Pi}_1$, $\bar{\Pi}_2$, $\bar{\Pi}_3$, and $\bar{\Pi}_4$ are mathematical expectations of ${\Pi}_1$, ${\Pi}_2$, ${\Pi}_3$, and ${\Pi}_4$ respectively. The compact representation of the closed-loop system is:
\begin{equation} \label{eq19}
\zeta(k+1)=\Gamma_1\zeta(k)+\Gamma_2\zeta(k)
\end{equation}
where
\begin{equation}
\begin{aligned}
{}&\zeta(k)=\begin{bmatrix} x(k) \\ e(k) \end{bmatrix} \\
&\Gamma_1= \begin{bmatrix} A+ B\Delta_2K & -B\Delta_2K \\0 & A-L\Delta_1C \end{bmatrix} \\
&\Gamma_2= \begin{bmatrix} B(\Delta_2-\bar{\Delta}_2)K & -B(\Delta_2-\bar{\Delta}_2)K \\ -L(\Delta_1-\bar{\Delta}_1)C & 0 \end{bmatrix}         \\
&\Delta_1={\Pi}_1+{\Pi}_2-{\Pi}_1{\Pi}_2 \\
&\Delta_2={\Pi}_3+{\Pi}_4-{\Pi}_3{\Pi}_4                            \nonumber
\end{aligned}
\end{equation}



\section{Controller Design} \label{s3}

In this section, the sufficient conditions for the stability of LTI systems under the attack  are derived. To deal with the stochastic variables defined in the attack models, it is necessary to use the concept of stochastic stability in the mean square sense. In the following, the notation of stochastic stability in the mean square sense is defined. 
\begin{definition}
The solution $\zeta=0$ for the closed-loop system given in (\ref{eq19}) is said exponentially stable in mean square sense if there exists constants $\rho \in [0,1]$ and $\sigma > 0$ such that
\begin{equation}
E\{\|\zeta(k)\|_2\}\leq \sigma \rho^kE\{\|\zeta(0)\|_2\}
\end{equation}
for any $\zeta(0)$ and any $k \geq 0$.
\end{definition}

Following Lemma is the base of deriving the sufficient conditions for mean square stability of the system given in (\ref{eq19}).
\begin{lemma}\cite{tarn1976observers}
If there exists real constant $\tau \in [0,1]$ such that
\begin{equation}
E\{V(\zeta(k+1)|\zeta(k))\}-V(\zeta(k)) \leq -\tau V(\zeta(k))
\end{equation}
and $V(k)$ is a quadratic Lyapunov function, then the sequence $\zeta(k)$ will be exponential mean square stable.
\label{lemma1}
\end{lemma}

\begin{theorem}
\label{t1}
The system given in (\ref{eq19}) is exponential mean square stable, if there exists matrices $Q_1 > 0$, $Q_2 > 0$, G, and H with appropriate dimensions such that satisfying the following LMI: 
\begin{equation}
\begin{bmatrix} -Q & \ast & \ast \\ \Sigma_1 & -Q & \ast \\ \Sigma_2 & 0 & -Q \end{bmatrix}
<0
\label{eq7}
\end{equation}
where
\begin{equation}
\begin{aligned} 
{} &Q=\begin{bmatrix} Q_1 & 0 \\ 0 & Q_2 \end{bmatrix} \\
&\Sigma_1= \begin{bmatrix} Q_1A+B\Delta_2G & -B\Delta_2G \\
0 & Q_2A-H\Delta_1C \end{bmatrix}             \\
&\Sigma_2=\begin{bmatrix}
B\Delta_{22}G & B\Delta_{22}G                                             \\
-H\Delta_{11}C & 0 \end{bmatrix}                                         \\
&\Delta_{11}=(Var\{\Delta_1\})^{1/2}                                   \\
&\Delta_{22}=(Var\{\Delta_2\})^{1/2}                     \nonumber
\end{aligned} 
\end{equation}
\end{theorem}

\textit{proof}: Consider the following Lyapunov fuction:
\begin{equation}
V(k)=x^{T}(k)Q_1x(k)+e^{T}(k)Q_2e(k)
\end{equation}
where $Q_1$ and $Q_2$ are the solution of (\ref{eq7}). Utilizing the results of lemma~\ref{lemma1}, we obtain:
\begin{equation}
\begin{aligned}
{} & E\{V((k+1)|V(k))\}-V(k)= \\ 
&E\{x^{T}(k+1)Q_1x(k+1)+e^{T}(k+1)Q_2e(k+1)| \\
&x(k),\ldots, x(0), e(k), \ldots, e(0)\} - x^{T}(k)Q_1x(k)- \\
&e^{T}(k)Q_2e(k)=E\{ [(A+B\Delta_2K)x(k)- B\Delta_2Ke(k)                                     \\ 
&+B(\Delta_2-\bar{\Delta}_2)Kx(k)-B(\Delta_2-\bar{\Delta}_2)Ke(k)]^T   \\
&Q_1[(A+B\Delta_2K)x(k)- B\Delta_2Ke(k)                                                    \\ 
&+B(\Delta_2-\bar{\Delta}_2)Kx(k)-B(\Delta_2-\bar{\Delta}_2)Ke(k)] \}         \\ 
&+E\{[(A-L\Delta_1C)e(k)-L(\Delta_1-\bar{\Delta}_1)Cx(k)]^T     \\
&Q_2[(A-L\Delta_1C)e(k)-L(\Delta_1-\bar{\Delta}_1)Cx(k)]\}\\
&-x^{T}(k)Q_1x(k)-e^{T}(k)Q_2e(k) 
\end{aligned}
\label{eq9}
\end{equation}

By taking mathematical expectation from (\ref{eq9}), we have:
\begin{equation}
\begin{aligned} 
{} & E\{V((k+1)|V(k))\}-V(k) \\
&= [(A+ B\bar{\Delta}_2K)x(k) -B\bar{\Delta}_2Ke(k)]^TQ_1 \\
&[(A+ B\bar{\Delta}_2K)x(k) -B\bar{\Delta}_2Ke(k)]+ \\
&[(A-L\Delta_1C )e(k)]^TQ_2 [(A-L\Delta_1C )e(k)]+ \\
&[B\Delta_{22}Kx(k)+B\Delta_{22}Ke(k) ]^T Q_1[B\Delta_{22}Kx(k)+ \\
&B\Delta_{22}Ke(k)]+[L\Delta_{11}Cx(k)]^TQ_2[L\Delta_{11}Cx(k)] \\
&-x^{T}(k)Q_1x(k)-e^{T}(k)Q_2e(k)=\zeta^T\Omega \zeta 
\end{aligned}
\label{eq10}
\end{equation}
where,
\begin{equation}
\begin{aligned} 
{} &\Omega= \begin{bmatrix} \Omega_1 \\ \Omega_2 \end{bmatrix}^T \begin{bmatrix} Q_1 & 0 \\ 0 & Q_2 \end{bmatrix} \begin{bmatrix} \Omega_1 \\ \Omega_2 \end{bmatrix} \\
&+\begin{bmatrix} B\Delta_{22}K & B\Delta_{22}K \\ L\Delta_{11}C & 0 \end{bmatrix}^T \begin{bmatrix} Q_1 & 0 \\ 0 & Q_2 \end{bmatrix}
\times \begin{bmatrix}  B\Delta_{22}K & B\Delta_{22}K \\ L\Delta_{11}C & 0  \end{bmatrix}\\
&- \begin{bmatrix} Q_1 & 0 \\ 0 & Q_2 \end{bmatrix}  
\end{aligned}
\label{eq10}
\end{equation}
\begin{equation}
\begin{aligned} 
{} &\Omega_1= \begin{bmatrix} A+B\bar{\Delta}_2K & -B\bar{\Delta}_2K \end{bmatrix} \\
&\Omega_2= \begin{bmatrix} 0 & A-L\bar{\Delta}_1C \end{bmatrix}        \\
&\Delta_{11}=[Var\{\Delta_1\}]^{1/2}, \Delta_{22}=[Var\{\Delta_2\}]^{1/2}
\end{aligned}
\end{equation}
\begin{figure*}[!b]
\centering
\normalsize
\hrulefill
\vspace*{4pt}
\begin{equation}
\label{eq11}
\begin{bmatrix} -Q_1 & \ast & \ast & \ast & \ast & \ast \\ 
0 & -Q_2 & \ast & \ast & \ast & \ast \\
A+B\Delta_2K & -B\Delta_2K & -Q_1^{-1} & \ast & \ast & \ast \\
0 & A-L\Delta_1C & 0 & -Q_2^{-1} & \ast & \ast \\
B\Delta_{22}K & B\Delta_{22}K & 0 & 0 & -Q_1^{-1} & \ast \\
L\Delta_{11}C & 0 & 0 & 0 & 0 & -Q_2^{-1} \end{bmatrix} 
< 0
\end{equation}


\end{figure*}

\begin{figure*}[!b]
\centering
\normalsize
\hrulefill
\vspace*{4pt}
\begin{equation}
\label{eq12}
\begin{bmatrix} -Q_1 & \ast & \ast & \ast & \ast & \ast \\ 
0 & -Q_2 & \ast & \ast & \ast & \ast \\
Q_1(A+B\Delta_2K) & -Q_1B\Delta_2K & -Q_1 & \ast & \ast & \ast \\
0 & Q_2(A-L\Delta_1C) & 0 & -Q_2 & \ast & \ast \\
Q_1B\Delta_{22}K & Q_1B\Delta_{22}K & 0 & 0 & -Q_1 & \ast \\
Q_2L\Delta_{11}C & 0 & 0 & 0 & 0 & -Q_2 \end{bmatrix} 
< 0
\end{equation}


\end{figure*}


If matrix $\Omega$ is negative definite, it is concluded that the closed-loop system given in (\ref{eq19}) is exponentially stable in mean square sense based on the lemma (\ref{lemma1}). Utilizing the Schur complement, the matrix $\Omega$ in (\ref{eq10}) is negative definite if (\ref{eq11}) is satisfied. thereafter, It is not hard to show that LMI (\ref{eq12}) implies (\ref{eq11}). The proof of the theorem (\ref{t1}) is completed by leveraging lemma (\ref{lemma2}). 

\begin{lemma}\cite{ho2003robust}
For the full rank matrix $B \in R^{m \times n }$ in the singular value decomposition form, there exist a nonsingular matrix $W$ such that $BW=Q_1B$, if and only if there exists a symmetric matrix $Q_1$ in the following form:
\begin{eqnarray}
Q_1&=& U \begin{bmatrix} Q_{11} & 0 \\ 0 & Q_{22} \end{bmatrix} U^T \label{eq13}\\
B&=&U \begin{bmatrix} B_0 \\ 0 \end{bmatrix} V^T
\end{eqnarray}
where $Q_{11} \in R^{m \times m} > 0$, $Q_{22} \in R^{(n-m)\times (n-m)} > 0$, $U \in R^{m \times m}$ and $V \in R^{n \times n}$ are unitary matrices, $B_0 \in R^{m \times m}$ is a diagonal matrix with positive diagonal elements.
\label{lemma2}
\end{lemma}
It follows from Lemma (\ref{lemma2}) that we can substitute $BW$ with $Q_1B$ in LMI (\ref{eq11}). Thereafter, we can easily see that the proof of theorem (\ref{t1}) is complete, if $G$ and $H$ are defined as following in (\ref{eq12})
\begin{eqnarray}
G&=&WK \label{eq14}\\
H&=&Q_2L.\label{eq15}
\end{eqnarray}

Solution to the LMI given in theorem (\ref{t1}) is utilized  to obtain controller gain ($K$) and observer gain ($L$) for mean square stability of the system defined in (\ref{eq19}). Assume $Q_1$, $Q_2$, $G$, and $H$ are the solution of (\ref{eq7}), and $Q_1$ is in the form of (\ref{eq13}). The matrix $W$ in lemma (\ref{lemma2}) is computed as:
\begin{equation}
U \begin{bmatrix} B_0 \\ 0 \end{bmatrix} V^TW=U \begin{bmatrix} Q_{11} & 0 \\ 0 & Q_{22} \end{bmatrix} U^T U \begin{bmatrix} B_0 \\ 0 \end{bmatrix} V^T
\end{equation}
which implies:
\begin{equation}
\begin{bmatrix} B_0 \\ 0 \end{bmatrix} V^TW=\begin{bmatrix} Q_{11} & 0 \\ 0 & Q_{22} \end{bmatrix} \begin{bmatrix} B_0 \\ 0 \end{bmatrix} V^T
\end{equation}

\begin{equation}
W=(B_0V^T)^{-1}Q_{11}B_0V^T.
\label{eq16}
\end{equation}

It is concluded from (\ref{eq14}), (\ref{eq15}), and (\ref{eq16}) that
\begin{eqnarray}
K&=&W^{-1}G \label{eq17}\\
L&=&Q_2^{-1}H \label{eq18}
\end{eqnarray}
The $K$ and $L$ calculated by (\ref{eq17}) and (\ref{eq18}) guarantee the exponential mean square stability of the LTI system (\ref{eq1}) when the sensors and actuators are under the attacks.



\section{Simulation Results} \label{s4}
In this section, the following third order LTI system under attacks is considered. 
\begin{equation}
\begin{aligned} 
{}&x(k+1)=\begin{bmatrix} -1.7 & -0.5 & 0.1 \\ 1 & 0 & -0.7 \\ 0 & 0.8 & 0 \end{bmatrix} x(k) + \begin{bmatrix} 1 & 0 \\ 0 & 1 \\ 0 & 0 \end{bmatrix} u(k) \\
&y(k)=\begin{bmatrix} \alpha_1 & 0 \\ 0 & \alpha_2 \end{bmatrix} \begin{bmatrix} 1 & 0 & 1\\ 0 & 1 & 0 \end{bmatrix} x(k)+(\begin{bmatrix} 1 & 0 \\ 0 & 1 \end{bmatrix}-\begin{bmatrix} \alpha_1 & 0 \\ 0 & \alpha_2 \end{bmatrix}) \\
&\times \begin{bmatrix} \beta_1 & 0 \\ 0 & \beta_2 \end{bmatrix} \begin{bmatrix} 1 & 0 & 1\\ 0 & 1 & 0 \end{bmatrix}x(k)
\end{aligned}
\label{eq20}
\end{equation}

Given (\ref{eq20}),  the observer dynamics and the controller attack model are:
\begin{equation}
\begin{aligned} 
{}&\hat{x}(k+1)=\begin{bmatrix} -1.7 & -0.5 & 0.1 \\ 1 & 0 & -0.7 \\ 0 & 0.8 & 0 \end{bmatrix}\hat{x}(k)+\begin{bmatrix} 1 & 0 \\ 0 & 1 \\ 0 & 0 \end{bmatrix}u(k) \\
&-L[y(k)-\begin{bmatrix} \bar{\alpha}_1 & 0 \\ 0 & \bar{\alpha}_2 \end{bmatrix}\begin{bmatrix} 1 & 0 & 1\\ 0 & 1 & 0 \end{bmatrix} \hat{x}(k) -(\begin{bmatrix} 1 & 0 \\ 0 & 1 \end{bmatrix}-\begin{bmatrix} \bar{\alpha}_1 & 0 \\ 0 & \bar{\alpha}_2 \end{bmatrix})\\
& \times\begin{bmatrix} \bar{\beta}_1 & 0 \\ 0 & \bar{\beta}_2 \end{bmatrix}\begin{bmatrix} 1 & 0 & 1\\ 0 & 1 & 0 \end{bmatrix})\hat{x}(k) \\
&u(k)=\begin{bmatrix} \gamma_1 & 0 \\ 0 & \gamma_2\end{bmatrix} K\hat{x}(k)+(\begin{bmatrix} 1 & 0 \\ 0 & 1 \end{bmatrix}-\begin{bmatrix} \gamma_1 & 0 \\ 0 & \gamma_2 \end{bmatrix} \\
&\times \begin{bmatrix} \delta_1 & 0 \\ 0 & \delta_2 \end{bmatrix})Kx(k)
\end{aligned}
\label{eq22}
\end{equation}
where

\begin{equation}
\begin{aligned} 
{}&E\{ \alpha_1\}=0.7, \ E\{ \alpha_2\}=0.8  \\
&E\{ \gamma_1\}=0.8, \ E\{ \gamma_2\}=0.9 
\end{aligned}
\label{eq22_1}
\end{equation}

The expected values of $\alpha_i(i=1,2)$ and $\gamma_i(i=1,2)$ show the fraction of time that sensors and actuators work normally. Given (\ref{eq22_1}), the sensors of the system are under attacks for 30 and 20 percent of the time in average respectively, and the actuators of the system are under attack for 20 and 10 percent of the time in average respectively. 

$\bar{\beta_i}(i=1,2)$ and $\bar{\delta_i}(i=1,2)$ for the simulation purpose is considered as follows:
$$
\bar{\delta_1}=\bar{\beta_1}=1.3, 
\bar{\delta_2}=\bar{\beta_2}=1.1
$$

\begin{figure}[!t]
\centering
\includegraphics[width=0.5\textwidth]{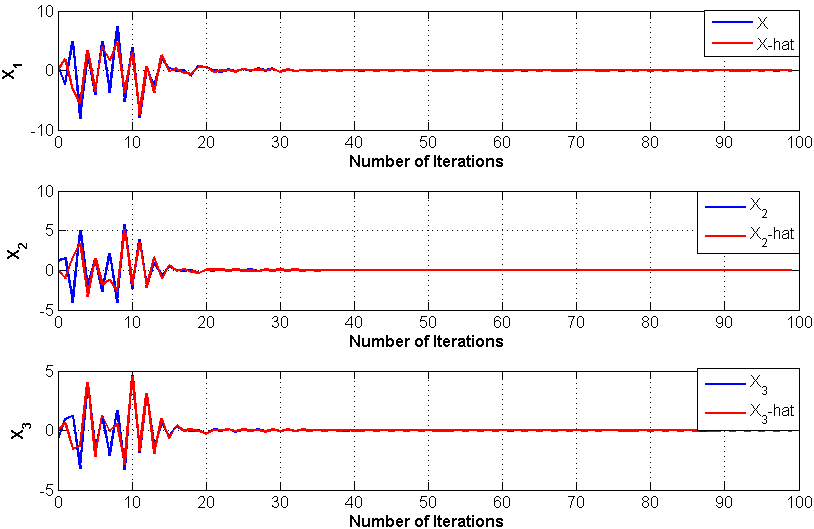}
\caption{The system states and observer states response under the attacks.}
\label{fig2}
\end{figure}

\begin{figure}[!t]
\centering
\includegraphics[width=0.5\textwidth]{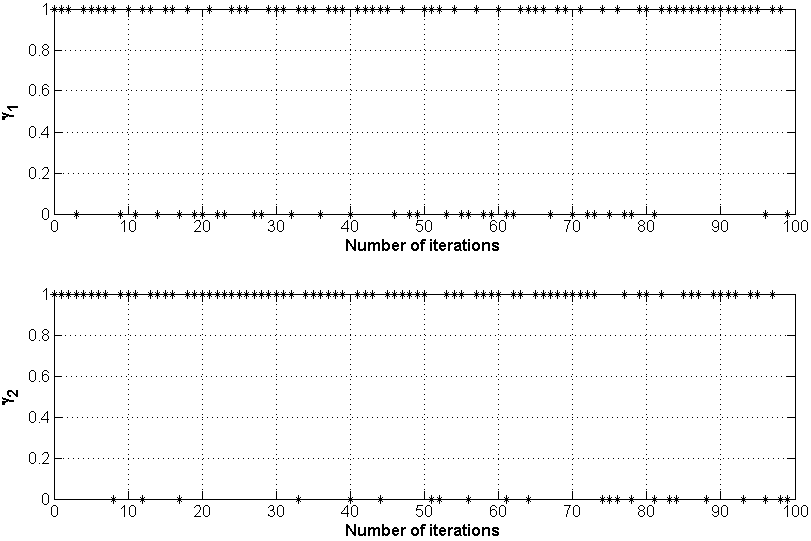}
\caption{The attack patterns on actuators of the system.}
\label{fig3}
\end{figure}

\begin{figure}[!t]
\centering
\includegraphics[width=0.5\textwidth]{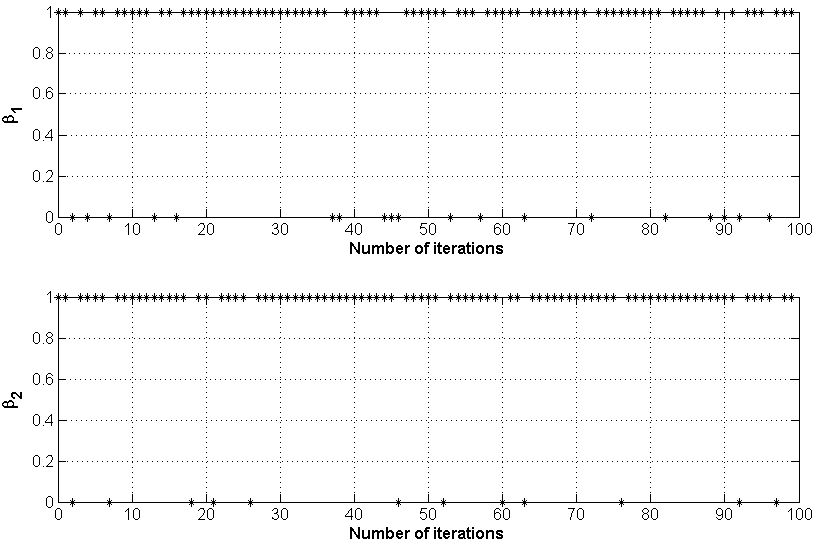}
\caption{The attack patterns on sensors of the system.}
\label{fig6}
\end{figure}

\begin{figure}[!t]
\centering
\includegraphics[width=0.5\textwidth]{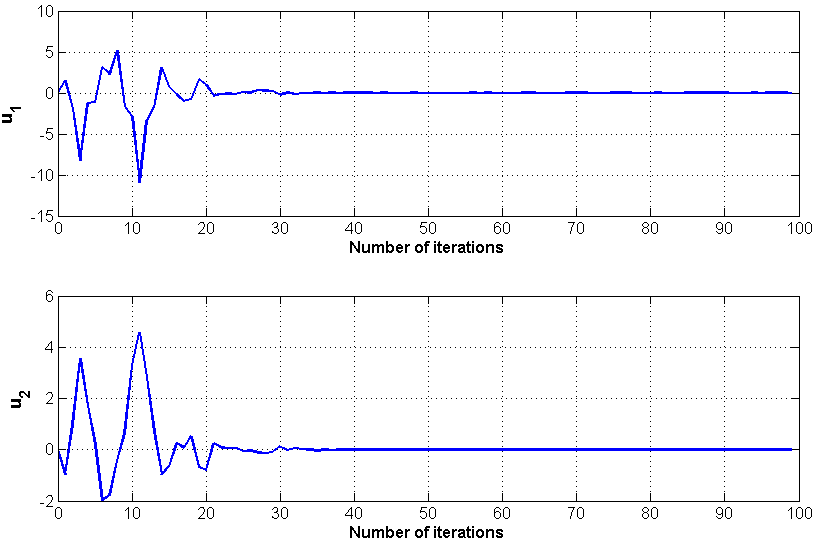}
\caption{The control signals of the actuators under attacks.}
\label{fig4}
\end{figure}

\begin{figure}[!t]
\centering
\includegraphics[width=0.5\textwidth]{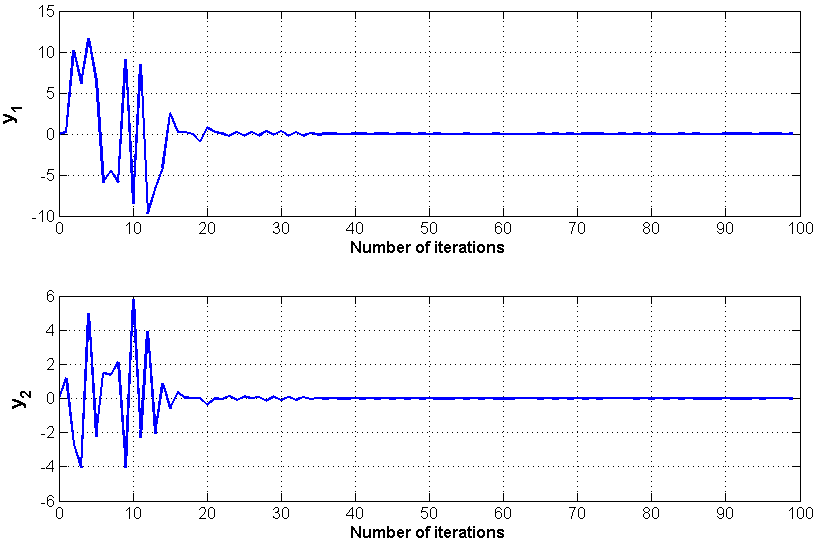}
\caption{The sensor measurement signals under attacks.}
\label{fig5}
\end{figure}

The solution to the LMI (\ref{eq7}) provides the feedback gain and the observer gain for the unstable system defined in (\ref{eq20}).
\begin{equation}
\begin{aligned} 
{}&K=\begin{bmatrix} 1.1475  &  -0.1962  & -1.4460\\ -0.7689  &  0.3120  &  1.3376 \end{bmatrix} \\
&L= \begin{bmatrix} 0.0674  &  1.5850 \\ -0.0376  & -0.8844\\ 0.0217  &  0.5095\end{bmatrix}.
\end{aligned}
\label{eq24}
\end{equation}

Given the initial conditions $x(0)=\begin{bmatrix} 1 & 1.2 & -0.8 \end{bmatrix} ^T$ and $\hat{x}(0)=\begin{bmatrix} 0 & 0 & 0 \end{bmatrix} ^T$, the state responses is shown in figure~\ref{fig2}. Figures \ref{fig3} and \ref{fig6} show the attack patterns on actuators and sensors of the system where $\alpha_i=0$ and $\gamma_1=0$ denote the system is under the attack. The actuator signals and the sensor measurements under the attacks are shown in figures \ref{fig4} and \ref{fig5} respectively. 



\section{Conclusion}
In this paper, a novel probabilistic model for the attack on sensors and actuators of LTI systems has been introduced. Denial-of-Service (Dos) attack is considered as a subclass of the introduced stochastic attack models. The probabilistic attacks on sensors and actuators are allowed to happen simultaneously in the system and the mean square stability of the system under attacks on sensors and the actuators is proved by leveraging stochastic control theory. The sufficient conditions for the existence of resilient controller gain and the observer gain are obtained, and it is shown that the problem of designing an output feedback controller is solvable if a certain LMI condition is satisfied.

\bibliographystyle{IEEEtran}
\bibliography{hossein}

\begin{thebibliography}{10}
\providecommand{\url}[1]{#1}
\csname url@samestyle\endcsname
\providecommand{\newblock}{\relax}
\providecommand{\bibinfo}[2]{#2}
\providecommand{\BIBentrySTDinterwordspacing}{\spaceskip=0pt\relax}
\providecommand{\BIBentryALTinterwordstretchfactor}{4}
\providecommand{\BIBentryALTinterwordspacing}{\spaceskip=\fontdimen2\font plus
\BIBentryALTinterwordstretchfactor\fontdimen3\font minus
  \fontdimen4\font\relax}
\providecommand{\BIBforeignlanguage}[2]{{%
\expandafter\ifx\csname l@#1\endcsname\relax
\typeout{** WARNING: IEEEtran.bst: No hyphenation pattern has been}%
\typeout{** loaded for the language `#1'. Using the pattern for}%
\typeout{** the default language instead.}%
\else
\language=\csname l@#1\endcsname
\fi
#2}}
\providecommand{\BIBdecl}{\relax}
\BIBdecl

\bibitem{keliris2016}
A.~Keliris, H.~Salehghaffari, B.~Cairl, P.~Krishnamurthy, M.~Maniatakos, and
  F.~Khorrami, ``Machine learning-based defense against process-aware attacks
  on industrial control systems,'' in \emph{Proceedings of the IEEE
  International Test Conference (ITC)}, TX, USA, Nov. 2016, pp. 1--10.

\bibitem{khodaparastan20121}
M.~Khodaparastan, H.~Vahedi, H.~Abyaneh, S.~Fathi, and G.~Gharehpetian,
  ``Impact of nonlinear controllers on dg's islanding phenomenon.''

\bibitem{super}
M.~Khodaparastan and A.Mohamed, ``Supercapacitors for electric rail transit
  systems,'' in \emph{Proceedings of the IEEE 6th International Conference on
  Renewable Energy Research and Applications (ICRERA)}, Nov. 2017, pp.
  896--901.

\bibitem{khodaparastan2017study}
M.~Khodaparastan and A.~Mohamed, ``A study on super capacitor wayside
  connection for energy recuperation in electric rail systems,'' in
  \emph{Proceedings of the Power \& Energy Society General Meeting}, July 2017,
  pp. 1--5.

\bibitem{ji2011resilient}
K.~Ji and D.~Wei, ``Resilient control for wireless networked control systems,''
  \emph{International Journal of Control, Automation and Systems}, vol.~9,
  no.~2, pp. 285--293, 2011.

\bibitem{rieger2012agent}
C.~Rieger, Q.~Zhu, and T.~Basar, ``Agent-based cyber control strategy design
  for resilient control systems: Concepts, architecture and methodologies,'' in
  \emph{Resilient Control Systems (ISRCS), 2012 5th International Symposium
  on}.\hskip 1em plus 0.5em minus 0.4em\relax IEEE, 2012, pp. 40--47.

\bibitem{giorgi2012adaptive}
S.~Giorgi, F.~Saleheen, F.~Ferrese, and C.-H. Won, ``Adaptive neural
  replication and resilient control despite malicious attacks,'' in
  \emph{Resilient Control Systems (ISRCS), 2012 5th International Symposium
  on}.\hskip 1em plus 0.5em minus 0.4em\relax IEEE, 2012, pp. 112--117.

\bibitem{rieger2010notional}
C.~G. Rieger, ``Notional examples and benchmark aspects of a resilient control
  system,'' in \emph{Resilient Control Systems (ISRCS), 2010 3rd International
  Symposium on}.\hskip 1em plus 0.5em minus 0.4em\relax IEEE, 2010, pp. 64--71.

\bibitem{rieger2009resilient}
C.~G. Rieger, D.~I. Gertman, and M.~A. McQueen, ``Resilient control systems:
  next generation design research,'' in \emph{Human System Interactions, 2009.
  HSI'09. 2nd Conference on}.\hskip 1em plus 0.5em minus 0.4em\relax IEEE,
  2009, pp. 632--636.

\bibitem{wei2010resilient}
D.~Wei and K.~Ji, ``Resilient industrial control system (rics): Concepts,
  formulation, metrics, and insights,'' in \emph{Resilient Control Systems
  (ISRCS), 2010 3rd International Symposium on}.\hskip 1em plus 0.5em minus
  0.4em\relax IEEE, 2010, pp. 15--22.

\bibitem{salehghaffari2018}
H.~Salehghaffari and F.~Khorrami, ``Resilient power grid state estimation under
  false data injection attacks,'' in \emph{Proceedings of the Innovative Smart
  Grid Technologies Conference (ISGT)}, Feb. 2018, pp. 1--5.

\bibitem{khodaparastan2017novel}
M.~Khodaparastan, H.~Vahedi, F.~Khazaeli, and H.~Oraee, ``A novel hybrid
  islanding detection method for inverter-based dgs using sfs and rocof,''
  \emph{IEEE Transactions on Power Delivery}, vol.~32, no.~5, pp. 2162--2170,
  Oct. 2017.

\bibitem{khodaparastan2012}
M.~Khodaparastan, A.~Mobarake, G.~Gharehpetian, and S.~Fathi, ``Smart fault
  classification in hvdc system based on optimal probabilistic neural
  networks,'' in \emph{Proceedings of the 2nd Iranian Conference on Smart Grids
  (ICSG)}, May 2012, pp. 1--4.

\bibitem{zhu2012dynamic}
Q.~Zhu and T.~Ba{\c{s}}ar, ``A dynamic game-theoretic approach to resilient
  control system design for cascading failures,'' in \emph{Proceedings of the
  1st international conference on High Confidence Networked Systems}.\hskip 1em
  plus 0.5em minus 0.4em\relax ACM, 2012, pp. 41--46.

\bibitem{hossein2017}
H.~Salehghaffari and F.~Khorrami, ``A game theoretic approach to design a
  resilient controller for a nonlinear discrete system,'' in \emph{Proceedings
  of the IFAC World Congress (to appear)}, Toulouse, France, July 2017.

\bibitem{elbsat2013robust}
M.~N. ElBsat and E.~E. Yaz, ``Robust and resilient finite-time bounded control
  of discrete-time uncertain nonlinear systems,'' \emph{Automatica}, vol.~49,
  no.~7, pp. 2292--2296, 2013.

\bibitem{li2015h}
Z.~Li, Z.~Wang, D.~Ding, and H.~Shu, ``H infinity fault estimation with
  randomly occurring uncertainties, quantization effects and successive packet
  dropouts: The finite-horizon case,'' \emph{International Journal of Robust
  and Nonlinear Control}, vol.~25, no.~15, pp. 2671--2686, 2015.

\bibitem{melin2013mathematical}
A.~M. Melin, E.~M. Ferragut, J.~A. Laska, D.~L. Fugate, and R.~Kisner, ``A
  mathematical framework for the analysis of cyber-resilient control systems,''
  in \emph{Resilient Control Systems (ISRCS), 2013 6th International Symposium
  on}.\hskip 1em plus 0.5em minus 0.4em\relax IEEE, 2013, pp. 13--18.

\bibitem{pasqualetti2007distributed}
F.~Pasqualetti, A.~Bicchi, and F.~Bullo, ``Distributed intrusion detection for
  secure consensus computations,'' in \emph{Decision and Control, 2007 46th
  IEEE Conference on}.\hskip 1em plus 0.5em minus 0.4em\relax IEEE, 2007, pp.
  5594--5599.

\bibitem{mo2009secure}
Y.~Mo and B.~Sinopoli, ``Secure control against replay attacks,'' in
  \emph{Communication, Control, and Computing, 2009. Allerton 2009. 47th Annual
  Allerton Conference on}.\hskip 1em plus 0.5em minus 0.4em\relax IEEE, 2009,
  pp. 911--918.

\bibitem{bezzo2014attack}
N.~Bezzo, J.~Weimer, M.~Pajic, O.~Sokolsky, G.~J. Pappas, and I.~Lee, ``Attack
  resilient state estimation for autonomous robotic systems,'' in
  \emph{Intelligent Robots and Systems (IROS 2014), 2014 IEEE/RSJ International
  Conference on}.\hskip 1em plus 0.5em minus 0.4em\relax IEEE, 2014, pp.
  3692--3698.

\bibitem{fawzi2014secure}
H.~Fawzi, P.~Tabuada, and S.~Diggavi, ``Secure estimation and control for
  cyber-physical systems under adversarial attacks,'' \emph{Automatic Control,
  IEEE Transactions on}, vol.~59, no.~6, pp. 1454--1467, 2014.

\bibitem{alpcan2004game}
T.~Alpcan and T.~Basar, ``A game theoretic analysis of intrusion detection in
  access control systems,'' in \emph{Proceedings of 43rd IEEE Conference on
  Decision and Control(CDC)}, vol.~2.\hskip 1em plus 0.5em minus 0.4em\relax
  IEEE, 2004, pp. 1568--1573.

\bibitem{tarn1976observers}
T.-J. Tarn and Y.~Rasis, ``Observers for nonlinear stochastic systems,''
  \emph{Automatic Control, IEEE Transactions on}, vol.~21, no.~4, pp. 441--448,
  1976.

\bibitem{ho2003robust}
D.~W. Ho and G.~Lu, ``Robust stabilization for a class of discrete-time
  non-linear systems via output feedback: the unified lmi approach,''
  \emph{International Journal of Control}, vol.~76, no.~2, pp. 105--115, 2003.

\end{thebibliography}

\end{document}